\documentclass[10pt,conference]{IEEEtran}
\usepackage{amsmath,graphicx}
\begin{document}

\newcommand{\ket}[1]{|#1\rangle}

\title{Conference Key Agreement and Quantum Sharing of Classical Secrets with
Noisy GHZ States}
\author{\authorblockN{Kai Chen and Hoi-Kwong Lo}
\authorblockA{
Center for Quantum Information and Quantum Control (CQIQC)\\
Department of Electrical and Computer Engineering (ECE) and\\
Department of Physics, University of Toronto, Toronto, ON M5S 3G4, Canada\\
Email: hklo@comm.utoronto.ca}
}

%


\maketitle

\begin{abstract}
We propose a wide class of distillation schemes for multi-partite
entangled states that are CSS-states. Our proposal provides not
only superior efficiency, but also new insights on the connection
between CSS-states and bipartite graph states. We then consider
the applications of our distillation schemes for two cryptographic
tasks---namely, (a) conference key agreement and (b) quantum
sharing of classical secrets. In particular, we construct
``prepare-and-measure'' protocols. Also we study the yield of
those protocols and the threshold value of the fidelity above
which the protocols can function securely. Surprisingly, our
protocols will function securely even when the initial state does
not violate the standard Bell-inequalities for GHZ states.
Experimental realization involving only {\it bi}-partite
entanglement is also suggested.
\end{abstract}

\section{Introduction}

Entanglement is the hallmark of quantum mechanics and has become
the most important resources for various quantum information tasks
nowadays. For practical aims, one may at first sight ask, how many
standard states (e.g., Einstein-Podolsky-Rosen (EPR) states and
Greenberger-Horne-Zeilinger (GHZ) states) can be distilled through
a noisy channel \cite{BDSW}. While this problem is really hard to
work out, except for the case of bipartite pure state, because one
has to consider all the possible strategy for entanglement
distillation (for a recent review, see \cite{Horo-qic}). A
practical way is to study specific strategies for entanglement
distillation.  While the classification of bipartite pure-state
entanglement has been solved, currently the classification of
multi-partite entanglement and mixed state bipartite entanglement
is an important open problem \cite{Horo-qic}. Our focus on this
paper is to study the distillation of {\it multi}-partite
entanglement and its applications to multi-party quantum
cryptography. Our motivations is threefold: theoretically for
better understanding and quantifying multi-partite entanglement;
practically to propose new applications of quantum cryptography in
the multi-party setting particularly in the presence of noises;
and finally to provide a bridge between theory and practice.

For convenience, we use the standard stabilizer formulation,
particularly specialize to the case where the output state is a
so-called Calderbank-Shor Steane (CSS) state \cite{CSS1,CSS2},
which is a simultaneous eigenstate of a complete set of
(commuting) stabilizer generators each of either X-type or Z-type.
We note that CSS states are equivalent to bipartite graph codes,
which have previously been analyzed by D\"{u}r, Aschauer and
Briegel \cite{graph}. We remark that one part of this
equivalence---that CSS states are bipartite graph states---was due
to Eric Rains \cite{rains}. Our observation puts the earlier work
of \cite{graph} in the more systematic setting of CSS formulation.
Moreover, we apply the idea of CSS-state distillation to the
construction of multi-party quantum cryptographic protocols.

We specifically consider two multi-party cryptographic tasks,
namely (a) conference key agreement where three parties, Alice,
Bob, and Charlie, would like to obtain a common random string of number,
known as the conference key, $k$, and to ensure that $k$ is secure
from any eavesdropper, Eve; and (b) quantum sharing of classical
secrets \cite{hbb99} where Alice would like to divide up her secret password
between two parties, B' and C', in such a way that neither B' nor
C' alone knows anything about the password and yet when B' and C'
come together, they can re-generate the password. To secure the
classical communications between the parties in both protocols, we
assume that each pair of the three parties are authenticated by
standard unconditionally secure method of authentication.

For practical interest, we will construct ``prepare-and-measure''
type protocols for both quantum cryptography tasks, for which the
participants do not need to have full-blown quantum computers to
implement them. In such protocols, there is a preparer
who prepares some number, say N, copies of
a standard entangled multi-partite state and distributes
them to other participants. Each of
other participants just performs some local individual measurement
on his/her share of each copy of the state. The participants then perform
classical post-processing (i.e.,
classical computations and classical communications
(CCCCs) that can be performed by strictly classical devices). Our
study is a natural generalization of the security proofs
by Shor-Preskill \cite{Shor-Preskill} and Gottesman-Lo
\cite{GLIEEE03} of the BB84 \cite{BB84}
quantum key distribution protocol
 to the multi-partite case. [The first proof of
security of BB84 was by Mayers \cite{mayersqkd}.]

Our paper is organized as follows. In Section~2, we study the entanglement
distillation of the GHZ state and present an improved hashing protocol.
It can distill prefect GHZ state successfully
whenever the fidelity $F \geq 0.7554$ comparing the result
of $F \geq 0.8075$ in \cite{smolincm2002} for a tripartite Werner-like state.
In Section~3, we generalize our results from the
GHZ state to a general so-called CSS state and show that the various subroutines
that we have studied, in fact, apply to a general CSS state. Also
we show the equivalence of CSS states and bipartite graph states.
In Sections~4 \& 5, we apply our formulation to study the three-party
conference key agreement and secret sharing problem, and show that
for tripartite Werner
state, conference key agreement is possible whenever the fidelity $F\geq
0.3976$, while for secret sharing whenever fidelity $F \geq 0.5372$.
We remarked that our protocols, will work even when the initial
GHZ-state does not violate standard Bell inequalities.

\section{Distillation of the GHZ state}

Suppose three distant parties, Alice, Bob, and Charlie, share
a GHZ state
$\left\vert \Psi \right\rangle _{ABC}=\frac{1}{\sqrt{2}}(\left\vert
0\right\rangle \left\vert 0\right\rangle \left\vert 0\right\rangle
+\left\vert 1\right\rangle \left\vert 1\right\rangle \left\vert
1\right\rangle$, which is the $+1$ eigenstate of commuting observables
(stabilizer generators):
$S_{0} =X\otimes X\otimes X, S_{1} =Z\otimes Z\otimes I,
S_{2} =Z\otimes I\otimes Z$, where $X=\begin{pmatrix}
0 & 1 \\
1 & 0%
\end{pmatrix}$,
$Y=\begin{pmatrix}
0 & -i \\
i & 0%
\end{pmatrix}
$,
$Z=\begin{pmatrix}
1 & 0 \\
0 & -1%
\end{pmatrix}$.
Let us denote a GHZ basis as following:
\begin{eqnarray}
\left\vert \Psi_{p,i_{1},i_{2}}\right\rangle _{ABC}=\frac{1}{\sqrt{2}}%
(\left\vert 0\right\rangle \left\vert i_{1}\right\rangle \left\vert
i_{2}\right\rangle +(-1)^{p}\left\vert 1\right\rangle \left\vert \overline{%
i_{1}}\right\rangle \left\vert \overline{i_{2}}\right\rangle ,
\label{GHZbasis}
\end{eqnarray}
where $p$ and the $i$'s are zero or one and a bar over a bit value indicates
its logical negation. Here, $(p,i_{1},i_{2})$
correspond to the
eigenvalues of the 3 stabilizer generators $S_{0},S_{1},S_{2}$ by
correspondence relation:
eigenvalue $1 \longrightarrow$ label 0, and eigenvalue $-1 \longrightarrow$ label 1.
Thus, a density matrix which is diagonal in the GHZ-basis
would be: $\rho_{ABC}=$diagonal$\{p_{000},p_{100},p_{011},p_{111},p_{010},p_{110},
p_{001},p_{101}\}$. We can think of an GHZ-state to be in one of the eight
possible basis states.

\subsection{Error rate estimation and derivation of density matrix (GHZ-basis
diagonal)}

Suppose three parties share a general tri-partite density
matrix that is not necessarily diagonal in the
GHZ-basis. But according to \cite{dctprl99,dcpra00}, one can depolarize
a general 3-party density matrix by applying with the operator $XXX$,
followed by $ZZI$ and finally, $ZIZ$ with a probability $1/2$ separately.
The overall operation makes $\rho $ diagonal in the basis Eq. (\ref{GHZbasis})
without changing the diagonal coefficients.
For this reason, we will focus only on the diagonal elements of
a density matrix in the GHZ-basis.

In this paper, we often assume the three parties share $n$ trios of qubits.
Notice that there is no need to assume an i.i.d. (independent identical
distribution) for the $n$ trios.
Instead, we consider the most
general setting where those $n$ trios
can be fully entangled among themselves and perhaps
also with some additional ancillas. As noted in the
last paragraph, we consider only the $n$-GHZ-basis for the $n$ trios.
Each basis vector can be denoted by a sequence of $n$ objects each of
which takes one of the $2^3=8$ possible values.
In fact, following Gottesman-Lo \cite{GLIEEE03},
we argue that the protocols presented
in this paper will work well (with high probability) so long as the
``type'' of sequence is
known. Recall that, to encode the information about the type, we need only
consider the relative frequency of the eight basis states for the $n$ trios.
We can easily summarize such information by the marginal density matrix:
$\rho_{ABC}=$diagonal$\{p_{000},p_{100},p_{011},p_{111},p_{010},p_{110},
p_{001},p_{101}\}$.

Now the three parties can estimate the eight matrix elements of the
aforementioned marginal density matrix reliably
by using local operations and
classical communications (LOCCs) only.
They just measure along $X, Y, Z$ basis and
compare the results of their local measurements.
In fact, they will find that
the error pattern for the seven non-trivial group elements of
$XXX, ZZI, ZIZ, -YYX$, $IZZ, -YZY, -XYY, -XYY$. For example, the error rate for
$XXX$ is $p_{100}+p_{101}+p_{110}+p_{111}$, the similar error pattern for other
group elements. Since these error rates are linearly dependent on $p_{ijk}$ and
can be determined by local operations and classical communications (LOCCs),
the above equations relate the diagonal
matrix element of the density matrix, $\rho_{ABC}$ to experimental observables.

It should be clear that the above error estimation procedure will give
accurate information on the type of the sequence. Therefore,
in what follows, when one studies
multi-partite entanglement distillation protocol, one can reduce
the problem of a general initial state of $n$ trios of qubits to the case where
the $n$ trios are treated as i.i.d. This reduction idea via the method of the type
is a generalization of the bipartite case studied
in \cite{GLIEEE03}. Such a reduction technique greatly simplifies the problem.

\subsection{Improved multi-party hashing for
distillation of GHZ states}

Suppose $M$ ($> 2$) parties share an ensemble of $n$
identical mixed multi-partite
states and they would like to distill out almost perfect (generalized)
GHZ states. Maneva and Smolin \cite{smolincm2002} found an
efficient multi-partite
entanglement distillation protocol---multi-party hashing method. They used
multilateral quantum XOR gates (MXOR) as shown in the Fig.4 of \cite{smolincm2002}
where every party imposes identical Control-NOT operations on some of their own
particles from one single source particle or to one single target particle.
If we write an unknown $N$-qubit
state as an $N$-bit string $b_0, b_1, b_2 , b_{N-1}$ where
$b_0$ corresponds to the eigenvalue of the operator
$XX \cdots X$ and $b_i$ ($i >0$) corresponds to the eigenvalue of the
operator $Z_1 Z_{i+1}$. (See Eq.~(3) of \cite{smolincm2002}.)
Note that $b_0$ denotes the phase
error pattern and $b_i$ ($i>0$) denotes the bit-flip
error pattern. Maneva and Smolin's protocol involves applying
first a number of rounds of random hashing in the amplitude bits followed
by performing a number of rounds of random hashing in
the phase bit(s). They showed that its yield (per input
mixed state) was, in the asymptotic large $n$ limit,
given by $D_{h}=1-\max_{j>0}[\{H(b_{j})\}]-H(b_{0})$,
where $H(x)$ is the standard Shannon entropy in classical information
theory \cite{Cover}. We will argue that, in fact, the yield can be increased to:
\begin{equation}
D_{h}^{^{\prime }}=1-\max
\{H(b_{1}),H(b_{2}|b_{1})\}-H(b_{0})+I(b_{0};b_{1},b_{2}),
\label{newyield}
\end{equation}
for tri-partite case where $I(X;Y)$ is the standard mutual information between X and Y.

The key point is that there may be some correlations between $b_{0}$ and $%
(b_{1},b_{2})$ and also between $b_{1}$ and $b_{2}$ i.e.,
$I(b_{0};b_{1},b_{2}) \geq 0, I(b_{1};b_{2}) \geq 0$
If these quantities are non-zero, one can improve Maneva-Smolin's protocol
by consider the following strategy:
\begin{enumerate}
\item[1] The three parties perform $n H(b_{1})$ rounds
of hashing to work out the value of $b_1$'s completely
(Suppose here $H(b_{1})\leq H(b_{2})$).
Afterwards, the uncertainty in the variables $b_{2}$'s is reduced to
$n H(b_{2}|b_{1})\}$. Therefore, only $n H(b_{2}|b_{1})\}$ rounds
of hashing is needed to work out the value of $b_2$'s.
Note that the hashing of $b_1$ and $b_2$ can be simultaneously executed.
Therefore, in total, we still only need $n [\max
\{H(b_{1}),H(b_{2}|b_{1})\}]$ rounds of random hashing in amplitude bits.

\item[2] They use the information on the pattern of $b_{1},b_{2}$ (the
amplitude bits) to reduce their ignorance on the pattern of $b_{0}$ (the phase bit)
from $n H(b_{0})$ to
$n H(b_{0}|(b_{1},b_{2}))=$ $n [ H(b_{0})-I(b_{0};b_{1},b_{2})] =n [ H(b_{0},b_{1},b_{2})-H(b_{1},b_{2})].$

\item[3]
They apply a random hashing by using multilateral quantum XOR gates with regard to
one single source particle of every party,
shown in detail in Fig.4b of \cite{smolincm2002} to identify the pattern of $b_{0}$.
Now only (slightly more than)
$n [H(b_{0},b_{1},b_{2})-H(b_{1},b_{2})] $ rounds of random hashing is needed.
\end{enumerate}
Thus the yield of our method gives Eq. (\ref{newyield}).

Suppose, one has prepared a class of Werner-like states
$\rho _{W}=\alpha \left\vert \Phi ^{+} \right\rangle \langle \Phi ^{+} |+\frac{%
1-\alpha }{2^{N}}I,\text{ \ }0\leq \alpha \leq 1,$
where $\ket{\Phi ^{+}} $ denotes the so-called cat state
and $I$ is the identity matrix. We remark that in the GHZ-basis, the state
can be rewritten into
$\rho _{W}=F\left\vert 0,00\ldots 0\right\rangle \langle 0,00\ldots 0| \nonumber \\
+\frac{1-F}{2^{N}-1}(I-\left\vert 0,00\ldots 0\right\rangle \langle 0,00\ldots 0|).
$
Using the random hashing method of
Maneva and Smolin, for the tri-partite case,
we can obtain perfect GHZ states with nonzero yield
whenever $F\geq 0.8075$. With our improved random hashing method, we still get
with nonzero yield whenever $F\geq 0.7554$. This is a substantial improvement of
the original method.

\section{CSS states}

In this section, we generalize our results on the GHZ state
to a general CSS state. We first define a CSS state
and show that, just like the GHZ state, a CSS state can be
distilled by the hashing protocol and a recurrence method developed by Murao et al
in \cite{mppvk98}. We derive the yield for the hashing protocol and
show that a CSS state is equivalent to a bipartite
(i.e., two-colorable) graph state, a subject of recent attention.

\subsection{Distillation of CSS-states}

A CSS-state is basically a CSS-code where the
number of encoded qubit is zero. For instance, an encoded $\ket{0}$ state
of a CSS code
is a CSS-state. More formally, we have the following definition.

{\bf CSS states:}
A CSS-state is a $+1$ eigenstate of a complete set of (commuting) stabilizer generators
such that each stabilizer element is of X-type or Z-type only. For example, a
GHZ-state is a CSS-state with stabilizer generators:
$XXX$, $ZZI$ and $ZIZ$.

{\bf Claim~1}:
Suppose we label its simultaneous eigenstate for a CSS-state by its simultaneous
eigenvalues $\ket{\hat{b} , \hat{p}}$ where $\hat{b}= \{b_1, b_2, \cdots,
b_m\}$
is a vector that denotes the tuples of Z-type eigenvalues and
$\hat{p}= \{p_1, p_2, \cdots, p_n\}$
is a vector that denotes the tuples of X-type eigenvalues.
Consider a pair of multi-partite states under multilateral
quantum XOR gates (MXOR), the state of the pair evolves as follows:
\begin{equation}
\text{MXOR} \left[ \ket{\hat{b^1} , \hat{p^1}} \ket{\hat{b^2} , \hat{p^2}}
\right] =
\ket{\hat{b^1} , \hat{p^1}+\hat{p^2} }\ket{\hat{b^1}+ \hat{b^2} , \hat{p^2}}.
\end{equation}

{\it Proof}: This follows from a standard result in quantum error correction.
It is easy to understand from the evolution of Pauli operators acting
on individual qubits.~Q.E.D.

It is easily shown that the yield is:
$D_{h_1} > 1 - \max_{i} \left[H(b_i)\right] - \max_{j} \left[H(p_j|\hat{b})\right]$.
This is so because the mutual information $I (\hat{b}, \hat{p})$ between
the bit-flip and phase syndrome is non-zero.
Alternatively, one can hash in X first and then hash in Z. Therefore,
we also have $D_{h_2} > 1 - \max_{j} \left[H(p_j)\right] - \max_{i}
\left[H(b_i|\hat{p})\right]$.
Combining the two results, we have $D_h \geq \max(D_{h_1},D_{h_2})$.

We note that one can apply the subroutines P1 and P2 in \cite{mppvk98}
to any CSS-state. The P1 and P2 steps appeared there are,
in fact, to apply bit-flip error detection (P2) and phase-flip
error detection (P1) separately. Here the
stabilizer for the GHZ state is CSS-like and errors can be corrected by
two steps together. Allow the successful distillation of
CSS-state at higher error rates than what is possible with only a
hashing protocol.

\subsection{Equivalence of CSS-states and bipartite graph states}

{\bf Bipartite graph state:}
A graph is said to be bipartite (i.e., 2-colorable) if the set of vertices can be
decomposed into two sets, say $L$ (left) and $R$ (right) such that
an edge is only allowed to connect a $L$ vertex with a $R$ vertex
(but not between two $L$ (or $R$) vertices).
A graph state is a multi-partite state where each vertex, $v_j$, represents
a qubit and has stabilizer generators of the form
$K_j = X_j \Pi_{(j,k) \in E} Z_k$.

Distillation of graph states, have
been recently discussed in \cite{graph} for a special class
of bipartite (i.e., 2-colorable). In what follows, we
will show that they are essentially equivalent to CSS-states.

{\bf Claim~2}: Bipartite (i.e, two-colorable)
graph states are equivalent to CSS-states.

\noindent {\it Proof}: For details, see our preprint \cite{GHZ} (one part of the
equivalence: CSS implies bipartite, is due to Eric Rains).

Claim~2 establishes the equivalence of two different mathematical
formulations: CSS-states and bipartite graph states.
Thus much of what we have learnt about the distillation of
bipartite/two-colorable graph states through the work of
\cite{graph} can be
interpreted in the more systematic language of CSS-states.
In particular, it is natural to consider the bit-flip and phase
error patterns separately and consider their propagations in
quantum computational circuit. From Claim~2, we learn that
Claim~1, which originally refers to CSS states, can be applied
directly to any bipartite graph states. Moreover, the
improvement that we have found in the last section, in fact,
applies to bipartite graph states.

\section{3-party conference key agreement with noisy GHZ states}

In this section, We consider a prepare-and-measure conference key
agreement scheme, where we allow the participants to perform only
local individual quantum measurements and local classical
computations and classical communications (CCCCs) to obtain the
same random string of secret, known as a key, which can be useful
for securing a conference call against eavesdropping attacks.

How do we prove the security of our protocol against an eavesdropper?
We start with a CSS-based GHZ state distillation protocol and try
to convert it to a ``prepare and measure" protocol.

\begin{figure}[tbh]
\resizebox{3.6cm}{!}{\includegraphics{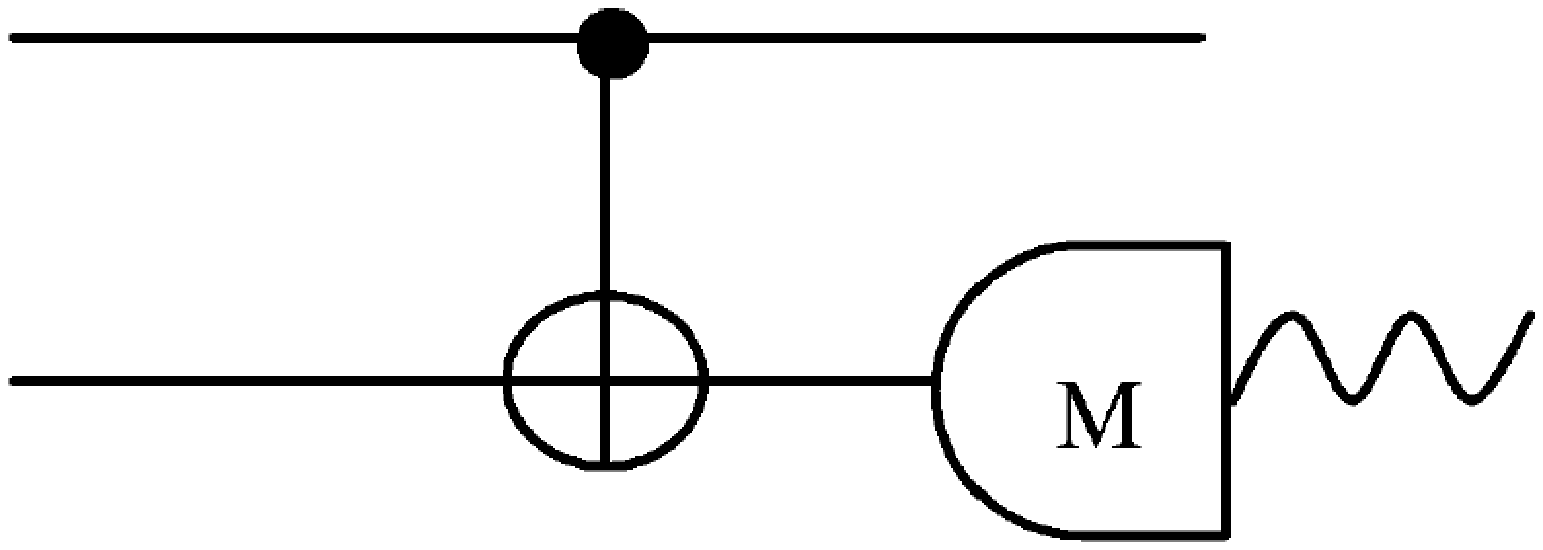}}
\hspace{0.5cm}%
\resizebox{4cm}{!}{\includegraphics{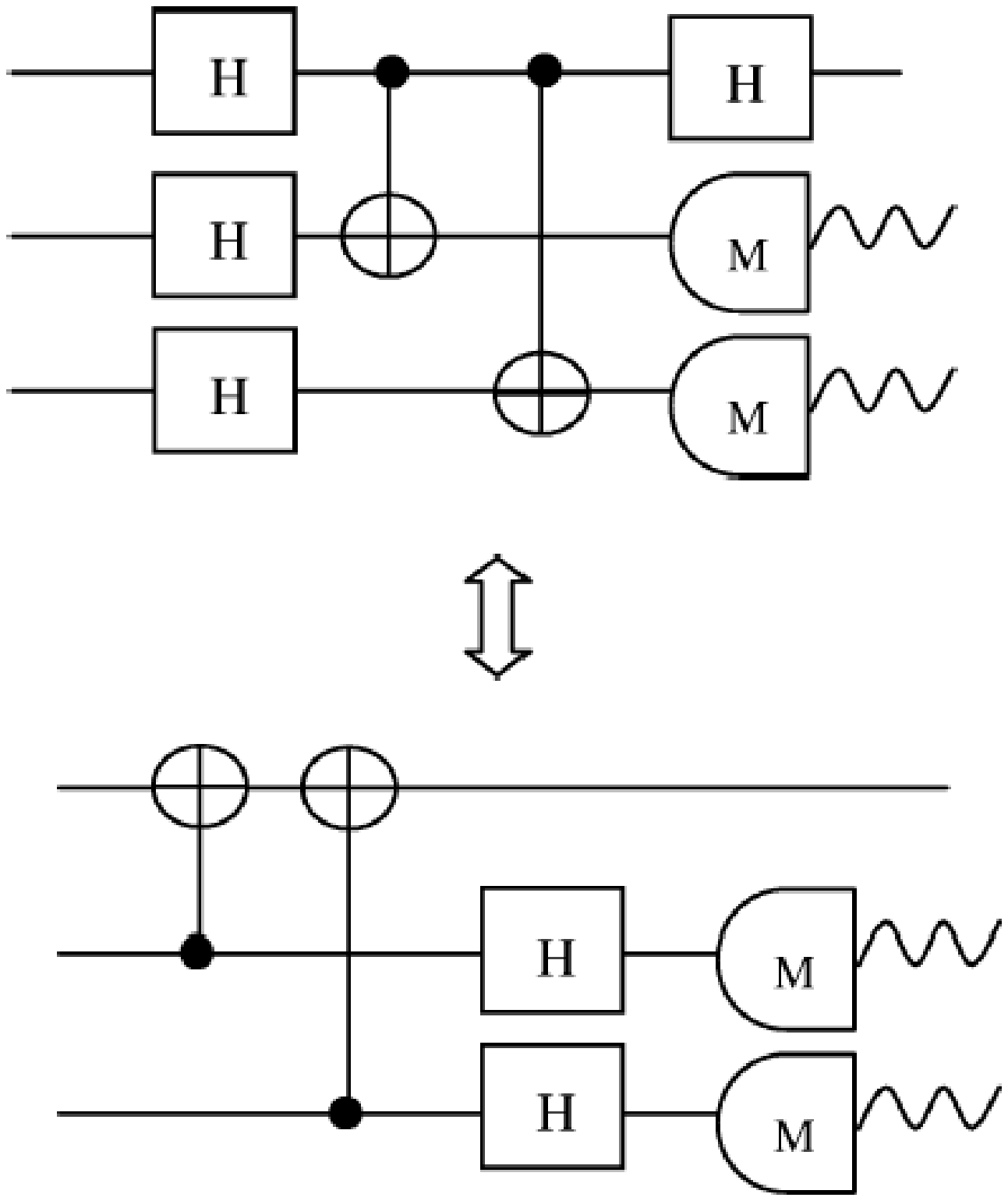}}
\caption{Left a: B step (bit-flip error detection) procedure for
conference key agreement protocol. Right b: P step (phase-flip error correction) procedure for conference
key agreement protocol. In a prepare-and-measure protocol
for conference key agreement, each of
Alice, Bob and Charlie simply takes the parity $(Z_{A}+Z_{B}+Z_{C})\mod 2$
of their own three particles, which can be done locally and no classical
communication is needed.}
\label{confkey}
\end{figure}

The protocol involves two subprotocols. For bit-flip error
detection (B step), we use the P2 step of
Murao et al \cite{mppvk98} as shown in Fig.~\ref{confkey}a. We remark
that the P1 step used in \cite{mppvk98}, is, in fact, a phase
error detection step (more precisely,
subsequent measurement operators applied are conditional to
phase error syndromes found in earlier measurements)
and is strictly forbidden in a
``prepare-and-measure" protocol \cite{GLIEEE03}. Thus for phase
error correction (P step), we design a procedure borrowing the
idea of Gottesman and Lo \cite{GLIEEE03} as shown in
Fig.~\ref{confkey}b, but apply it to the multi-partite case.

\textbf{Our multi-partite B step:} Using the
language of stabilizer, this step can be
reformulated by the following transformation of the two GHZ-like states%
\begin{eqnarray}
&&[(p,i_{1},i_{2}),(q,j_{1},j_{2})] \nonumber \\
&&\overset{\text{applying BXOR}}{\longrightarrow }(p\oplus
q,i_{1},i_{2}),(q,i_{1}\oplus j_{1},i_{2}\oplus j_{2}),
\end{eqnarray}%
where $(p,i_{1},i_{2}),(q,j_{1},j_{2})$ denote the phase bit and amplitude
bits for the first and second GHZ-like states. If $i_{1}\oplus
j_{1}=i_{2}\oplus j_{2}=0 \mod 2$, we keep the first GHZ state, otherwise discard
all the two states. This corresponds to the prescription that we keep the first trio iff $%
M_{A}=M_{B}$ and $M_{A}=M_{C}$ (i.e., Alice, Bob and Charlie get the same
measurement outcome). This step just changes the 8 elements of diagonal entries accordingly
and can be obtained by straight calculation.

\textbf{Our multi-partite P step:} This step (shown
in Fig.~\ref{confkey}b) can be reformulated by the following transformation of the three
GHZ-like states%
\begin{align}
&[(p,i_{1},i_{2}),(q,j_{1},j_{2}),(r,k_{1},k_{2})] \nonumber \\
&\overset{\text{applying 1st BXOR}}{\longrightarrow }[(p,i_{1}\oplus
j_{1},i_{2}\oplus j_{2}),(p\oplus q,j_{1},j_{2}),(r,k_{1},k_{2})] \nonumber \\
&\overset{\text{applying 2nd BXOR}}{\longrightarrow }  \nonumber \\
&[(p,i_{1}\oplus j_{1}\oplus k_{1},i_{2}\oplus j_{2}\oplus k_{2}),(p\oplus
q,j_{1},j_{2}),(p\oplus r,k_{1},k_{2})]
\end{align}%
If $p\oplus q=p\oplus r=1\mod 2$, we apply $p\longrightarrow p\oplus 1%
\mod 2$, otherwise keep the first GHZ-like state invariant. Note that P
can also be performed locally by each party, which regards the circuit as
implementing a 3-qubit phase error correction code.

We now argue that the aforementioned entanglement distillation subprotocol
(P step) can be converted to a prepare-and-measure protocol.
This is in the spirit of \cite{GLIEEE03}.
In conference key agreement, Alice, Bob and Charlie do not need to perform
phase error correction. They only need to prove that, phase
error correction would have been successful, if they had performed it.
Thus in the second part of Fig.~\ref{confkey}b, each of
Alice, Bob and Charlie simply takes the parity $(Z_{A}+Z_{B}+Z_{C})\mod 2$
of their own three particles. No classical
communication is needed. Moreover, we can apply the same conversion
idea to any concatenated
protocols involving B steps, P steps and following by a hashing protocol.
This conversion result means that we can obtain secure protocols
by considering the convergence
of GHZ distillation protocols involving those operations.

By direct numerical calculation, we can verify that our scheme can
distill GHZ state with nonzero yield whenever $F\geq 0.3976$ by
some state-dependent sequence of B and P steps, and then change to
our random hashing method if it works. [Notice that the yield of a
recurrence protocol asymptotically goes to zero
(see \cite{BDSW}). Therefore, it is advantageous to switch to
a hashing protocol at some point.] We find that a sequence of
B and P steps BBBBB for $F=0.3976$, which is optimal for any
sequence with at most 5 steps. our new protocol gives dramatic
improvement by using 2-way classical communications compared with
the random hashing method of Maneva and Smolin which works only
when $F\geq 0.8075$.

\section{Quantum sharing of classical secrets in a noisy channel}

Quantum sharing of classical secrets---has
also been introduced \cite{hbb99}.
Here, the goal is to use quantum states to share a {\it classical}
secret between multiple parties and to ensure that no eavesdropper
can learn useful information by passive eavesdropping.
Whereas \cite{hbb99} considers only the perfect case, here we
consider the case where the initial state is imperfect,
following \cite{Gisin1,Gisin2}.

{\em Phase One}:
The key point is that the GHZ state is a $+1$ eigenstate of $XXX$,
and satisfy a classical constraint $X_A + X_B + X_C = 0 \mod 2$.
Note that individually, each of Bob and Charlie has no information
on $X_A$. But, by getting together, Bob and Charlie can obtain
$X_B + X_C \mod 2$ and, therefore, obtain $X_A$.

{\em Phase two:} Suppose Alice now would like to share a bit value, $b$, with Bob and Charlie.
She can broadcast $X_A + b$.

\begin{figure}[tbh]
\resizebox{3.6cm}{!}{\includegraphics{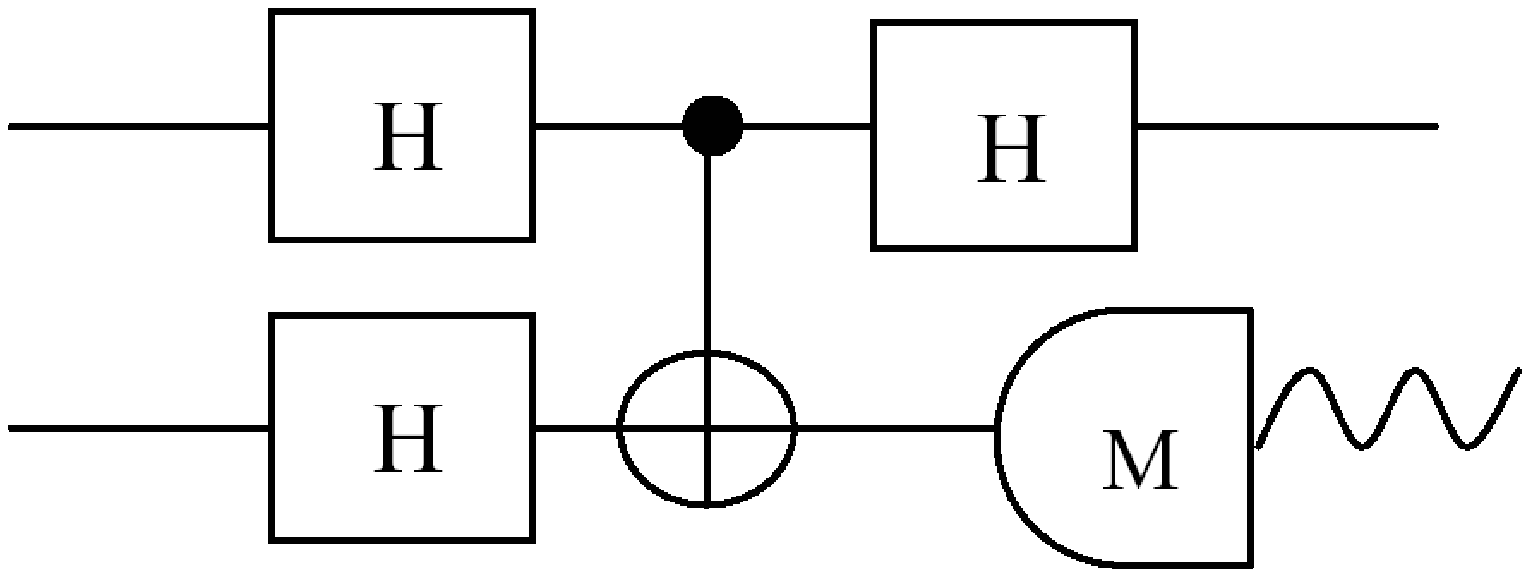}}
\hspace{0.5cm}%
\resizebox{4cm}{!}{\includegraphics{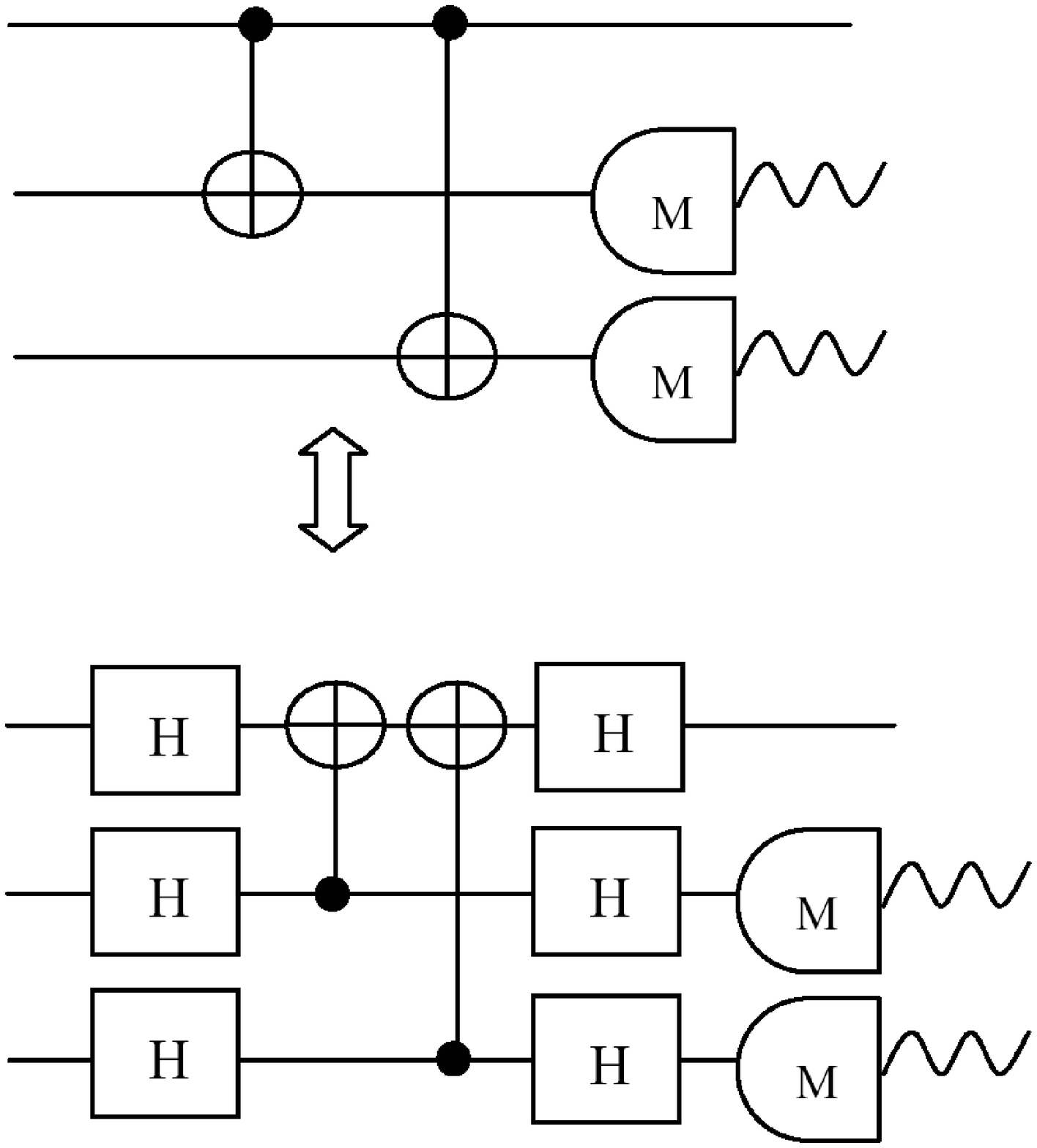}}
\caption{Left a: B' step (bit-flip error detection) procedure for secret sharing.
Right b: P' step (phase-flip error correction) procedure for secret sharing.
In a prepare-and-measure protocol for secret sharing, each of
Alice, Bob and Charlie simply takes the parity $(X_{A}+X_{B}+X_{C})\mod 2$
of their own three particles, which can be done locally and no classical
communication is needed.}
\label{ss}
\end{figure}

Note that, since the final measurement
is now done along the X-axis, the bit-flip measurements now correspond to
measurements along X.  Therefore, for the bit-flip error detection code (B'
step), we use the same procedure as the P1 step of Murao et al \cite{mppvk98}
as shown in Fig.~\ref{ss}a. As for the phase error correction (P' step), we design a
procedure similar to the idea of Gottesman and Lo \cite{GLIEEE03} as shown
in Fig.~\ref{ss}b. Details of our protocol can be found in our
preprint \cite{GHZ}.

Similar argument to conference key agreement case holds for secret sharing.
The three parties only need to ensure that, phase
error correction (P' step) would have been successful, if they had performed it.
We verify
that our scheme used in secret sharing in a noisy channel can distill GHZ
state with nonzero yield whenever $F\geq 0.5372$ by some state-dependent
sequence of B and P steps and then change to our random hashing method if it
works. For the optimal sequences within 5 steps, we find it is
(B'B'B'B'B') that just gives $F\geq 0.5372$ followed by hashing method.

{\bf Remark 1:} Note that, if one of the parties, say Alice,
is actually the preparer of the multi-partite state, in
a prepare-and-measure protocol, she is allowed to pre-measure
her subsystem. By doing so, she projects an $N$-partite
entangled state into one of the various $(N-1)$-partite entangled
state. Therefore, conference key agreement and secret sharing protocols
can be implemented with only
$(N-1)$-party entanglement.

{\bf Remark 2:}
It was claimed in Refs. \cite{Gisin1,Gisin2} that a violation of
Bell inequalities is a criterion for security of secret sharing
schemes \cite{hbb99} under the assumption of individual attacks
by Eve and {\it one}-way classical post-processing protocols
by Alice and Bob. Violation of Bell inequalities for
$N$ particles Werner-like state is shown in Ref. \cite{zb2002} to
be $\alpha >1/\sqrt{2^{(N-1)}}$. For a tripartite system, this
gives $\alpha >1/2$ and thus $F>9/16 \doteq 0.5625$. This is
clearly a higher requirement for the initial fidelity of a
Werner-like state than that for our two-way prepare and measure
secret sharing scheme which only requires $F\geq 0.5372$. Thus our
two-way protocols are secure even when Bell inequalities are not
violated.

\noindent {\bf Experimental Implementations.} From Remark~1,
for the three-party case, our protocols can be done
with only bi-partite entangled
states and can be experimentally implemented
with, for example, parametric down conversion sources.
More concretely, imagine that Alice prepares a perfect GHZ
state and measures her qubit along the $X,Y,Z$-axis. After her
measurement, Bob and Charlie's state will be $ { 1 \over \sqrt{2}
} ( \ket{00} + \ket{11} )$ or $ { 1 \over \sqrt{2} } ( \ket{00} -
\ket{11} )$; $ { 1 \over \sqrt{2} } (
\ket{00} +i \ket{11} )$ or $ { 1 \over \sqrt{2} } ( \ket{00} -
i\ket{11} )$; $\ket{00}$ or $\ket{11}$
(with equal probabilities) separately. Thus Alice could implement
the two protocols by simply preparing one of the six states,
and sending the two qubits to Bob and Charlie
respectively through some quantum channels.

We remark that our protocols are not proven to be
optimal. In future, it will be interesting to search for
protocols with better yields and higher threshold error
rates. Moreover, one may try to generalize our results to quantum sharing
of classical secrets for more general access structures. Our results
will shed some light on the fundamental questions of the
classification of multi-party entanglement, the power and
limitations of multi-party quantum cryptography.
The details of our work can be found in \cite{GHZ}.

We thank helpful discussions with many colleagues and financial
support from a number of funding agencies.

\end{document}